\documentclass[twoside,slac_one]{revtex4}
\usepackage{graphicx}
\usepackage{fancyhdr}
\usepackage{amsmath} 
\usepackage{bm}
\usepackage{amsxtra}
\usepackage{amssymb}
\usepackage{amsthm}
\usepackage{latexsym}
\usepackage{lscape}

\begin{document}

\title{Chiral and deconfinement transitions in lattice QCD with  improved staggered action}

%

\author{A. Bazavov and P. Petreczky for HotQCD Collaboration}
\affiliation{Physics Department,  Brookhaven National Laboratory, NY 11973 USA}

\begin{abstract}
We discuss chiral and deconfinement aspects of the finite temperature
transition in QCD using improved staggered actions. 
We study different quantities related to chiral and deconfinement transition
and discuss their cutoff dependence. Contrary to some
earlier lattice results we find that the chiral and deconfinement transition
are not closely interconnected.
\end{abstract}

\maketitle

\thispagestyle{fancy}


\section{Introduction}

QCD is expected to undergo a transition to a deconfined state, where thermodynamics
can no longer be described by hadronic degrees of freedom but should be described in terms of elementary
quark and gluon degrees of freedom. In addition, the chiral symmetry which is broken in
QCD vacuum is expected to be restored above some temperature. In the limit of zero quark masses
the restoration of the chiral symmetry is expected to be  a true phase transition. However,
for the quark masses realized in  nature this transition turns out to be an analytic crossover 
\cite{milc04thermo,nature,rbcbi06}. A question naturally arises whether the deconfinement and the chiral transitions
are closely related. Early lattice calculations with large quark masses and/or coarse lattices
suggested that deconfinement and chiral transition happen at the same temperature \cite{oldrev}.
However, more recent investigations that use so-called stout staggered quark action and finer lattices found
that these two transitions are no longer interconnected \cite{stoutTc06,stoutTc09,stoutTc10}. 
In this paper we are going to discuss the deconfinement and chiral transition in QCD at non-zero
temperature using highly improved staggered quark (HISQ) action and tree-level improved gauge
action. We refer to this combination of quark and gauge actions as HISQ/tree action. To control
discretization effects calculations have been performed at three values of the lattice
spacing corresponding to temporal extent $N_{\tau}=6,~8$ and $12$. To fix the lattice spacing
we used the $r_1$ scale of the static quark potential \cite{milc04} and the kaon decay constant $f_K$.
Additional calculations using the asqtad action with $N_{\tau}=8$ and $12$ have been performed to
demonstrate the consistency of the results obtained with different actions, since the asqtad action
was extensively used in the past to study QCD at non-zero temperature \cite{milc04thermo,milc06eos,hoteos}.

\section{Chiral transition}
\begin{figure}
\includegraphics[width=0.45\textwidth]{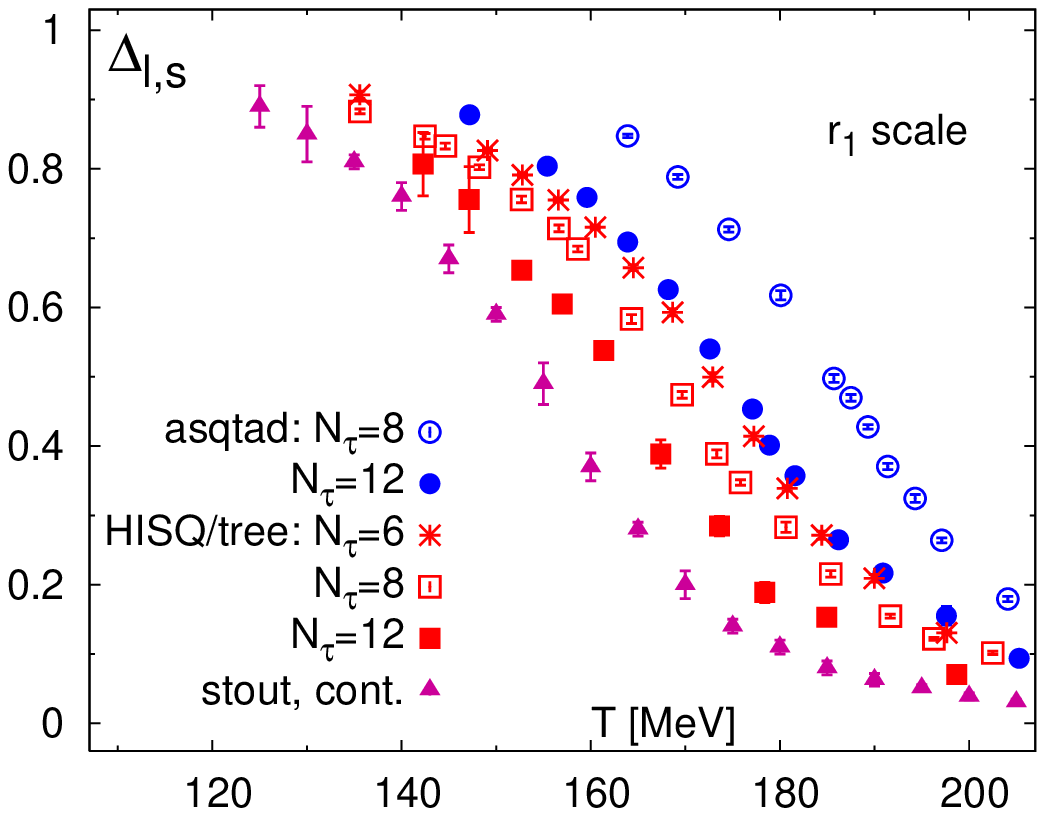}
\includegraphics[width=0.45\textwidth]{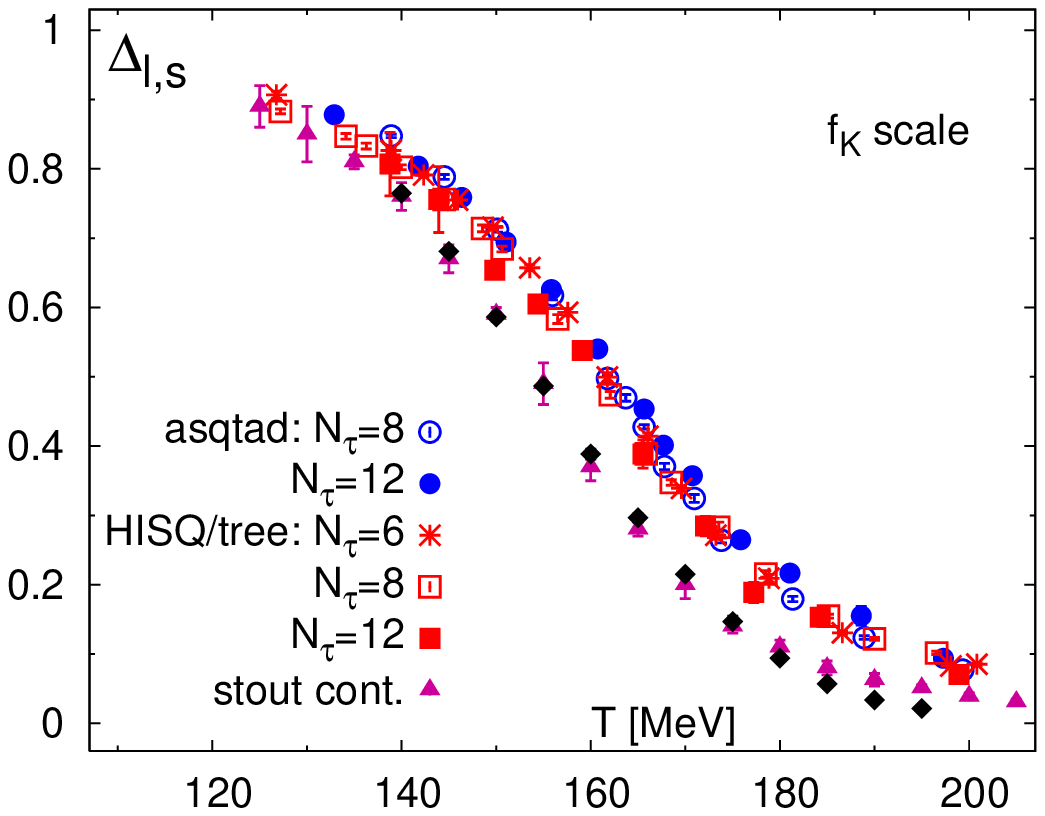}
\caption{The subtracted chiral condensate for the asqtad and HISQ/tree actions for $m_l=m_s/20$
is compared with the continuum extrapolated data obtained with the stout action \cite{stoutTc10} (left panel).
The temperature $T$ is converted into physical 
units using $r_1$ in the left panel.
In the right panel we show the temperature dependence of the same subtracted chiral condensate 
for the HISQ/tree and asqtad actions using $f_K$ to set the scale.
The black diamonds show HISQ/tree results for $N_{\tau}=8$ after interpolation to
the physical quark mass $m_l=0.037m_s$.
}
\label{fig:Deltals}
\end{figure}

The breaking of the chiral symmetry in QCD vacuum is signaled by non-zero expectation value of
quark condensate $\langle \bar \psi \psi \rangle$. At non-zero temperature the quark condensate is
expected to decrease, signaling the restoration of the chiral symmetry. However, the quark condensate needs
a multiplicative, and for non-zero quark mass, also an additive renormalization. Therefore following
Ref. \cite{rbcbi07} we consider the following quantity, which we will call the renormalized chiral
condensate
\begin{equation}
\Delta_{l,s}(T)=\frac{\langle \bar\psi \psi \rangle_{l,\tau}-\frac{m_l}{m_s} \langle \bar \psi \psi \rangle_{s,\tau}}
{\langle \bar \psi \psi \rangle_{l,0}-\frac{m_l}{m_s} \langle \bar \psi \psi \rangle_{s,0}}.
\end{equation}
Here $\langle \bar\psi \psi \rangle_{l,0}$ and  $\langle \bar\psi \psi \rangle_{l,\tau}$ refer to quark
condensate at zero and non-zero temperatures, $q=l$ and $s$ for light and strange quarks, respectively.
The numerical results are shown in Fig. \ref{fig:Deltals} using the lattice spacing determined by
$r_1$ parameter and $f_K$. We also show the continuum estimate for $\Delta_{l,s}$ 
obtained with the stout action in Fig. \ref{fig:Deltals}.
We use the value $r_1=0.3106$fm \cite{r1val} and $f_K=156.1$MeV \cite{fKval} when setting the scale in MeV.
When $r_1$ is used to set the scale we see large deviations for asqtad action, while for HISQ/tree action
these deviations are largely reduced. Interestingly enough, when $f_K$ is used to set the scale almost no
cutoff effect is seen in $\Delta_{l,s}$ both for HISQ/tree and asqtad action. This feature was first noticed for stout
action \cite{stoutTc09}.  The difference in the stout action and our result is due to the 
small difference in the light quark mass $m_l$. In our calculation $m_l=m_s/20$, 
while the stout calculations
correspond to $m_l=0.037m_s$. Here $m_s$ is the physical strange quark mass. If we perform interpolation in the
quark mass using $O(N)$ scaling, which can describe the quark mass dependence of the chiral observables obtained
with p4 action very well \cite{meos09,meos11}, to the value $m_l=0.037m_s$ we get a very good agreement with the
stout results.

\section{Deconfinement transition}

\begin{figure}
\includegraphics[width=0.49\textwidth]{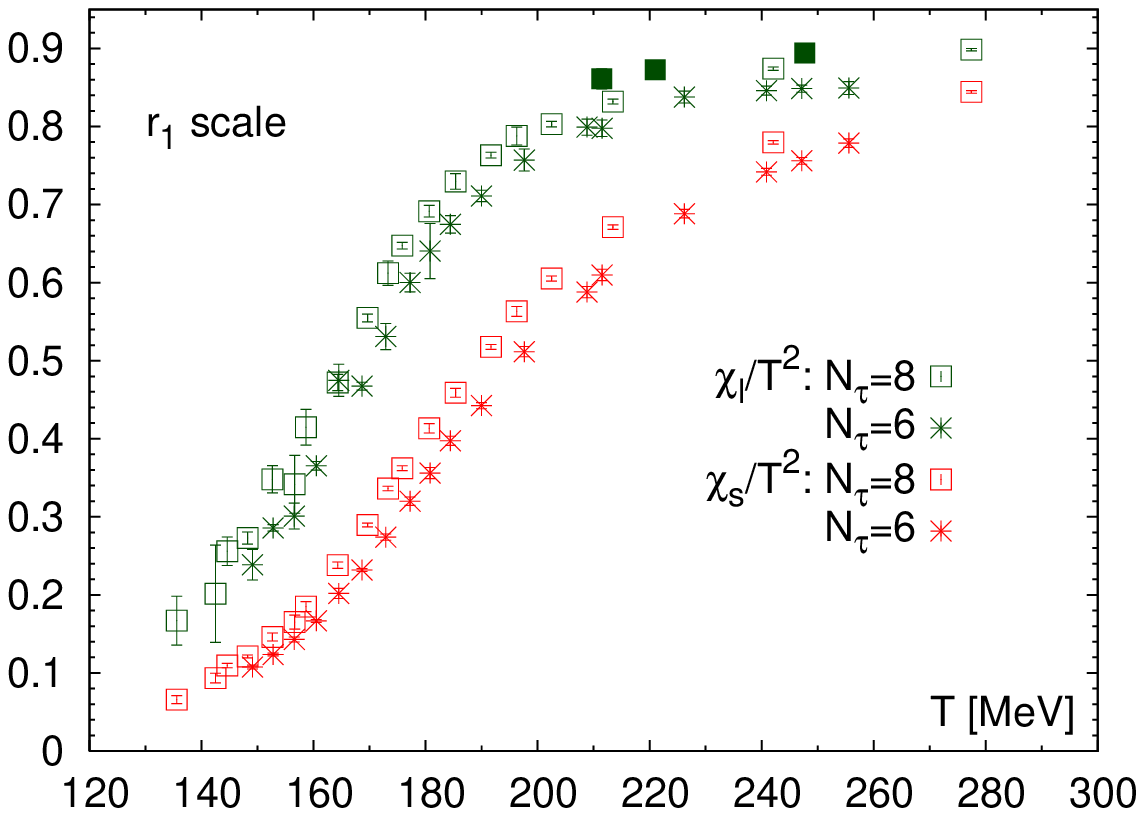}
\includegraphics[width=0.49\textwidth]{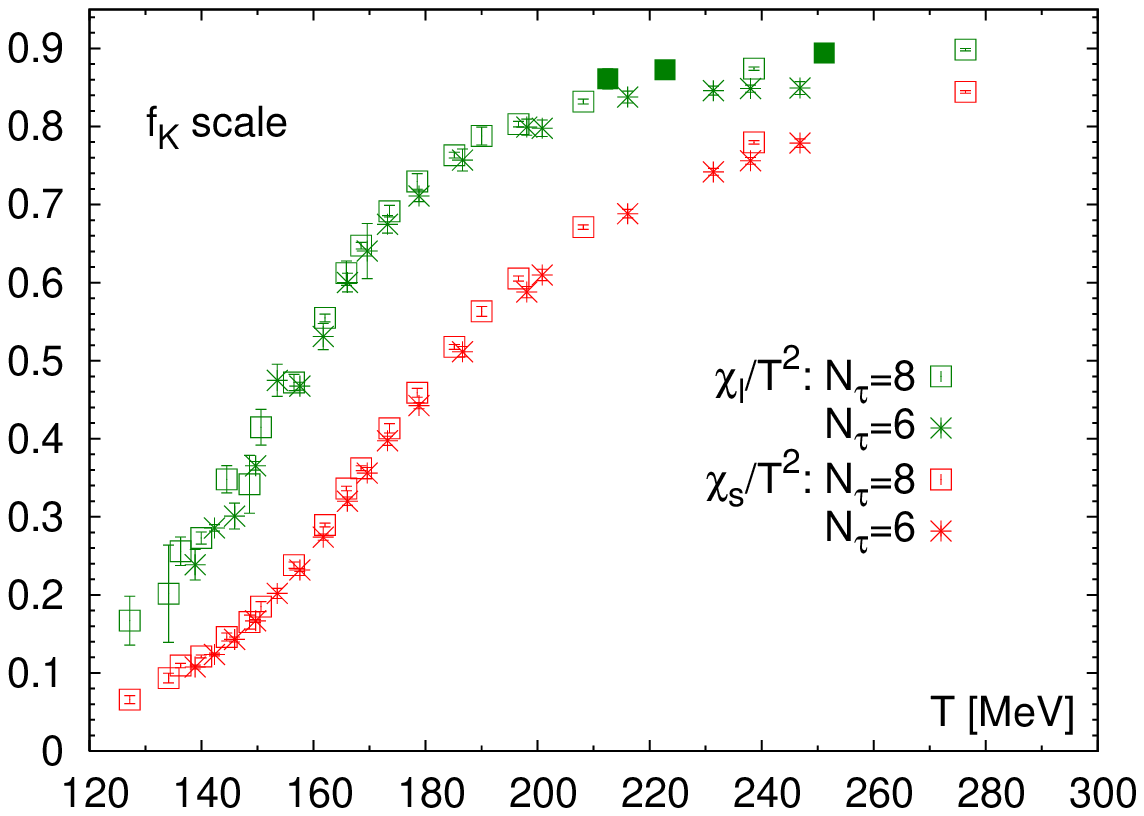}
\caption{Light quark number susceptibilities calculated for the asqtad and 
HISQ/tree actions and compared with the strange quark number susceptibility. In the 
left panel $r_1$ is used to set the lattice scale, while in the 
right panel we use $f_K$. The filled squares correspond to $N_{\tau}=12$.
}
\label{fig:chiq}
\end{figure}

Quark number susceptibilities, i.e. fluctuations of the quark numbers are sensitive probe
of deconfinement. These can be defined as second derivatives with respect to quark chemical
potential evaluated at zero chemical potentials
\begin{equation}
\chi_q=\frac{T}{V} \frac{\partial^2 \ln Z}{\partial \mu_q^2}|_{\mu_q=0},~~~ q=l,s.
\end{equation}
At low temperatures quark number fluctuations are determined by massive hadrons and therefore
are quite small, while at high
temperatures they are determined by light quark degrees of freedom and thus proportional to $T^2$.
The deconfinement transition can bee
seen as a rapid change between these two limiting behaviors and thus the quark number susceptibilities 
are expected to show a rapid increase.
In Fig. \ref{fig:chiq} we show the light and strange quark number susceptibilities and we clearly see
the expected rapid rise in these quantities. As before the lattice spacing was fixed using
$r_1$ and $f_K$. If $f_K$ is used to fix the scale cutoff effects turn out to be very small.
The rapid rise in the light quark number susceptibilities happens at temperatures, where $\Delta_{l,s}$
sharply decreases. The strange quark susceptibility shows a rapid rise at somewhat higher 
temperatures. Note, however, that this behavior of quark number susceptibilities is not related
to different transition temperatures. The inflection points of quark number susceptibilities are
dominated by the regular part of the free energy density, and the difference in the inflection points
is simply due to the difference in the quark mass.

\begin{figure}
\includegraphics[width=7.79cm]{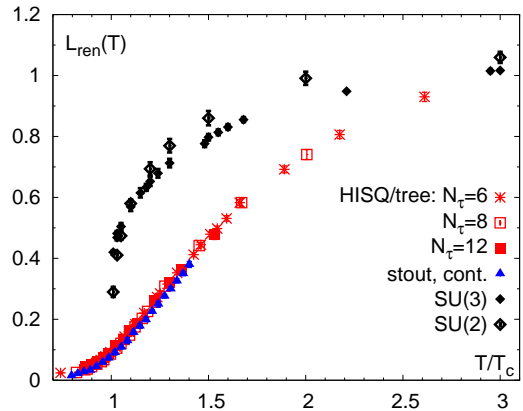}
\caption{
The renormalized Polyakov
loop as function of $T/T_c$ (right). For the HISQ/tree data we used the values of $T_c$ discussed in \cite{qm11},
while for stout data we used  the value of $T_c=157$ MeV from the inflection point of renormalized chiral condensate
\cite{stoutTc10}.}
\label{fig:Tc_and_Lren}
\end{figure}

The Polyakov loop is an order parameter for the deconfinement transition in pure gauge theory,
which is governed by $Z(N)$ symmetry. For QCD this symmetry is explicitly broken
by dynamical quarks. There is no obvious reason for the Polyakov loop
to be sensitive to the singular behavior close to the chiral limit although speculations
along these lines have been made \cite{speculations}. The Polyakov loop 
is related to the screening properties of the medium and thus to deconfinement.
After proper renormalization, the square of the Polyakov loop characterizes the
long distance behavior of the static quark anti-quark free energy; it 
gives the excess in free energy needed to screen two well-separated color
charges. The renormalized Polyakov loop has been studied in the past in 
pure gauge theory \cite{okacz02,digal03} as well as in QCD with two 
\cite{okacz05}, three \cite{kostya04} and  two plus one flavors 
\cite{rbcbi07,hoteos}. The renormalized Polyakov loop, calculated on lattices 
with temporal extent $N_\tau$, is obtained from the bare Polyakov loop
\begin{eqnarray}
&
\displaystyle
L_{ren}(T)=z(\beta)^{N_{\tau}} L_{bare}(\beta)=
z(\beta)^{N_{\tau}} \left<\frac{1}{3}  {\rm Tr } 
\prod_{x_0=0}^{N_{\tau}-1} U_0(x_0,\vec{x})\right >,
\end{eqnarray}
where $z(\beta)=\exp(-c(\beta)/2)$ and $c(\beta)$ is the additive normalization
of the static potential chosen such that it coincides with the string potential
at distance $r=1.5r_0$ with $r_0$ being the Sommer scale. This procedure 
of normalizing the Polyakov loop follows Ref. \cite{stoutTc06}. Some earlier
calculations used the singlet free energy in Coulomb gauge to estimate the renormalized
Polyakov loop \cite{okacz02,digal03,okacz05,kostya04}. While the former procedure 
is expected to be more precise both procedures give the same results within errors.
The numerical results for the renormalized Polyakov loop for the HISQ/tree action are
shown in the right panel of Fig.~\ref{fig:Tc_and_Lren} as function of $T/T_c$, with $T_c$
being the transition temperature. 
As one can see from the figure the cutoff 
($N_{\tau}$) dependence of the renormalized Polyakov loop is small. We also compare
our results with the continuum extrapolated stout results \cite{stoutTc10} and the 
corresponding results in pure gauge theory \cite{okacz02,digal03}. 
We find good agreement between our results and the stout results.
We also see that in the vicinity of the transition temperature the behavior of the renormalized
Polyakov loop in QCD and in the pure gauge theory is quite different.

\section{Conclusion}
In this contribution we discussed different quantities, which characterize the deconfinement and
chiral transition in QCD at non-zero temperature and studied their cutoff dependence.
We showed that when the kaon decay constant 
$f_K$ is used to set the scale (lattice spacing) the cutoff effects in different quantities are
quite small and calculations performed with the asqtad and HISQ/tree actions are in good agreement
with calculations performed with the stout action. We pointed out that it is difficult to
define the deconfinement temperature. Different observables that are used to characterize the deconfinement
transition show rapid rise at different temperatures, which in turn, could be different from the chiral transition
temperature. This is due to the fact that the observables used to study the deconfinement transition are not
sensitive to the singular part of the free energy density or have limited sensitivity to it.

\begin{acknowledgments}
This work has been supported in part by contracts DE-AC02-98CH10886
and DE-FC02-06ER-41439 with the U.S. Department of Energy
and contract 0555397 with the National Science Foundation. The numerical calculations have been performed
using the USQCD resources at Fermilab as well as the BlueGene/L
at the New York Center for Computational Sciences (NYCCS).
\end{acknowledgments}

\bigskip 

\end{document}